\documentclass[conference]{IEEEtran}
\IEEEoverridecommandlockouts
\usepackage{cite}
\usepackage{amsmath,amssymb,amsfonts}
\usepackage[ruled,linesnumbered]{algorithm2e}
\usepackage{graphicx}
\usepackage{subcaption}
\usepackage{textcomp}
\usepackage{xcolor}
\usepackage{booktabs}
\usepackage{multirow}
\usepackage{url}
\usepackage[table]{xcolor}
\definecolor{lightgray}{rgb}{0.9,0.9,0.9}
\def\BibTeX{{\rm B\kern-.05em{\sc i\kern-.025em b}\kern-.08em
    T\kern-.1667em\lower.7ex\hbox{E}\kern-.125emX}}
\begin{document}

\title{A high-capacity linguistic steganography based on entropy-driven rank-token mapping\\
}


\author{
\IEEEauthorblockN{Jun Jiang,
Weiming Zhang,
Nenghai Yu,
and Kejiang Chen\textsuperscript{*}
}
\IEEEauthorblockA{\textit{School of Cyber Science and Technology} \\
\textit{University of Science and Technology of China} \\
Hefei, Anhui, China \\
\{jungle0430@mail., chenkj@\}ustc.edu.cn}
}

\maketitle

\begin{abstract}
Linguistic steganography enables covert communication through embedding secret messages into innocuous texts; however, current methods face critical limitations in payload capacity and security. Traditional modification-based methods introduce detectable anomalies, while retrieval-based strategies suffer from low embedding capacity. Modern generative steganography leverages language models to generate natural stego text but struggles with limited entropy in token predictions, further constraining capacity. To address these issues, we propose an entropy-driven framework called \textit{RTMStega} that integrates rank-based adaptive coding and context-aware decompression with normalized entropy. By mapping secret messages to token probability ranks and dynamically adjusting sampling via context-aware entropy-based adjustments, \textit{RTMStega} achieves a balance between payload capacity and imperceptibility. Experiments across diverse datasets and models demonstrate that \textit{RTMStega} triples the payload capacity of mainstream generative steganography, reduces processing time by over 50\%, and maintains high text quality, offering a trustworthy solution for secure and efficient covert communication.
\end{abstract}

\begin{IEEEkeywords}
linguistic steganography, covert communication, payload capacity, rank-token mapping
\end{IEEEkeywords}

\section{Introduction}
Steganography is an important technique in the field of information hiding, which enables covert communication by embedding secret messages in innocent carriers, thus avoiding malicious monitoring by third parties~\cite{b1,b2,b4}. Unlike encrypted communications, which may attract suspicion due to their unnatural format, stego texts blend seamlessly into regular communication, offering an effective solution for privacy protection and secure covert communication~\cite{b5,b7}. Among the various communication media, text stands out as an ideal medium for steganography due to its particular prevalence in everyday communications (e.g., emails, social media posts, and instant messages). Therefore, this paper focuses on text steganography, which is also called linguistic steganography~\cite{b3,b6}.


The general process of linguistic steganography involves encoding a secret message into a format compatible with the carrier, referred to as the cover text. This involves embedding the message within the cover text, transmitting the resultant stego text via a public channel, and ultimately extracting and decoding the concealed message at the recipient's end~\cite{hopper2002provably, rani2013text, krishnan2017overview, majeed2021review}. Traditional linguistic steganography primarily utilizes modification-based strategies, e.g., synonym substitution~\cite{xiang2014linguistic}, text paraphrasing~\cite{chang2010linguistic}, and spelling transformation~\cite{shirali2008text}. However, these methods often suffer from limited embedding capacity and susceptibility to statistical detection due to heuristic modifications that may introduce detectable anomalies. In contrast, retrieval-based methods utilize large-scale text libraries to encode samples and select appropriate utterances for transmission based on the embedded secret messages~\cite{chen2015coverless, xiang2018linguistic, wang2019coverless}. This method benefits from conveying covert messages without altering the original text, thus making it undetectable. Nevertheless, it requires the prior sharing of a substantial corpus to prevent contextual discrepancies between the communicating parties and often struggles to achieve high payload capacity~\cite{wang2023hi}. The limited capacity of steganography necessitates multiple interactions between communicating parties, thereby increasing the risk of suspicion by third parties.

With advancements in generative models, steganography has found new avenues for innovation. Serving as an efficient sampler, the generative model can generate text that is close to the distribution of human-generated content through continuous iterative optimization~\cite{qwen2.5, 2025deepseekr1}. Crucially, these models can also provide explicit output probabilities for lexical elements. If secret messages can guide the model in producing specific content while maintaining or only slightly altering the sampling distribution, high-quality and imperceptible stego text can be generated. Leveraging these capabilities, generative steganography has emerged as a leading method in the field. It employs the probability distribution of the language model to map secret bit stream to token sequences, thereby synthesizing the natural stego text~\cite{chen2018provably,yang2018rnn,yang2020vae,zhang2020linguistic,zhou2021linguistic,yangpixel}.

Building upon this, steganography that leverages deep generative models~\cite{yang2018rnn,yang2020vae} and autoregressive models~\cite{zhang2020linguistic,zhou2021linguistic} has been developed, which encodes secret bit stream by segmenting the probability space of predicted tokens. Furthermore, frameworks that ensure provable security enhance the computational indistinguishability of the stego text from the cover text by meticulously adjusting their distributions~\cite{chen2018provably, yangpixel}. For instance, adaptive dynamic grouping (\textit{ADG})~\cite{ADG} dynamically groups tokens based on their sample probability and encodes messages according to the index of these groups. \textit{METEOR}~\cite{METEOR} introduces a range reversible sampling method that encodes messages as offsets within sampling intervals while simultaneously optimizing payload capacity through shared prefixes. \textit{Discop}~\cite{Discop} breaks down the process of high-dimensional token selection into multiple binary decisions by creating copies of the distribution and Huffman tree. Despite these advancements, generative steganography continues to encounter a significant challenge: low payload capacity. Although modern large language models (LLMs) produce outputs that are significantly more varied and fluid, the reduced entropy in their token predictions further limits their capacity.

To address these issues, we propose a linguistic steganography based on entropy-driven rank-token mapping called \textit{RTMStega}. Drawing inspiration from the text compression capabilities of LLMs and their context-adaptive compression encoding~\cite{2023llmzip}, our framework innovatively merges rank-based adaptive coding context-aware decompression with normalized entropy. We find that even when employing the same compression coding, the texts decompressed in various contexts display significant differences. Leveraging this characteristic, we embed secret messages by exploiting the differences between stego context and private context via compression coding. Specifically, secret messages are first compressed into a rank sequence, then transcoded into a context-adaptive $\beta$-bit representation, and finally reconstructed into coherent stego text under stego context through normalized-entropy-based decoding. This method, based on rank-token mapping, not only markedly enhances the payload capacity but also preserves the naturalness and contextual coherence of the stego text.

In summary, our contributions are detailed as follows:
\begin{itemize}
\item We analyze the risks associated with covert communication stemming from the limited payload capacity of current linguistic steganography. To address this, we propose a linguistic steganography based on entropy-driven rank-token mapping called \textit{RTMStega}.
\item We innovate by utilizing token probability ranks as an intermediate stego representation and adapt the sampling process based on the entropy of the model’s output sampling distribution. By adjusting relevant parameters, we achieve an effective trade-off between payload capacity and text quality, thereby enhancing the efficiency of covert communication.
\item Our experimental evaluations across multiple datasets and models demonstrate that \textit{RTMStega} triples the payload capacity and reduces steganographic processing time by over 50\% compared to baseline methods while maintaining comparable stego text quality.
\end{itemize}

\section{Related Work}
\subsection{Generative Language Model as Sampler.}
Modern generative language models generate text through iterative token sampling from explicit probability distributions using autoregressive mechanisms~\cite{qwen2.5, 2025deepseekr1}. At time step $t$, the token probability distribution is represented as $\mathbf{p}_t = (p_1, p_2, \dots, p_{|V|})$, where $|V|$ denotes the vocabulary size. The sampling operation transforms a random variable $u \in [0,1)$ into a token $w_t$ through:
\begin{equation}
    w_t = \arg\max_k \left\{ k \in \{1,\dots,|V|\} \,\bigg|\, \sum_{i=1}^{k-1} p_i \leq u < \sum_{i=1}^k p_i \right\}.
\end{equation}
This framework enables covert data embedding through the dynamic partitioning of the unit interval. For secret bit stream encoding $\mathbf{m} = (m_1, m_2, \dots, m_n)$, the $[0,1)$ interval is divided into token-specific sub-intervals proportionally aligned with their probabilities. The sub-interval containing $u$ simultaneously determines the generated token $w_t$ and the embedded secret bit stream.

\begin{figure*}[t]
  \includegraphics[width=\textwidth]{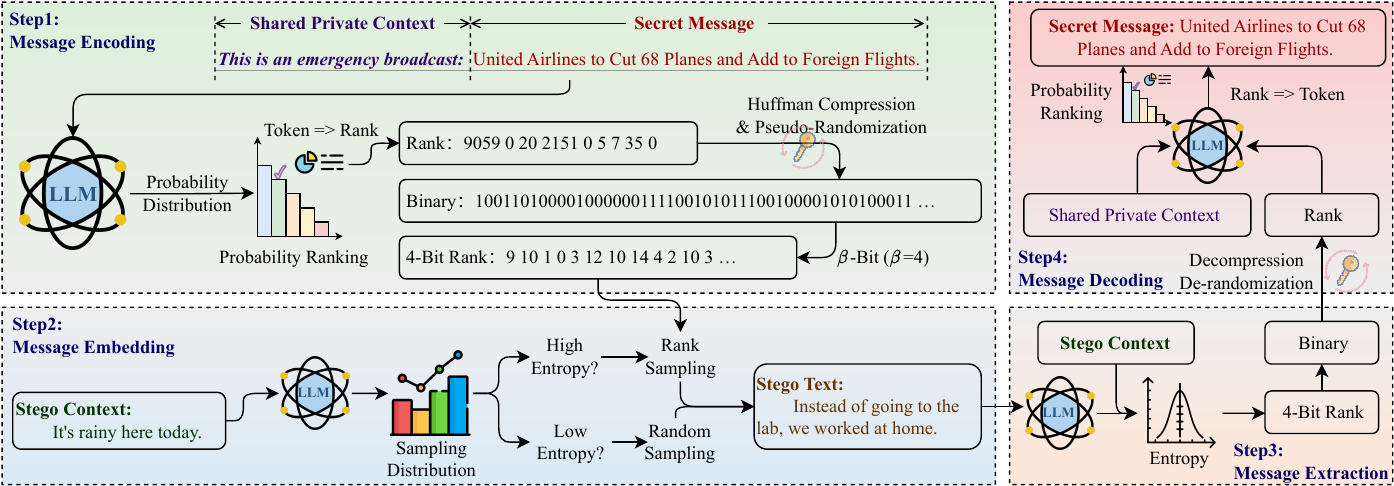}
  \caption{The overall framework of \textit{RTMStega}, consisting of four steps: message encoding, message embedding, message extraction, and message decoding.}
  \vspace{-0.4cm}
  \label{fig:framework}
\end{figure*}

\subsection{Generative Steganography.}
Recent advances in generative steganography utilize autoregressive language models to embed secret messages in generated stego text while maintaining statistical indistinguishability from cover text~\cite{yang2018rnn,yang2020vae,zhang2020linguistic,zhou2021linguistic,chen2018provably, yangpixel}. Given a language model $\mathcal{M}$ that generates tokens through conditional probability distribution $P(x_t \mid x_{<t}, \mathcal{H})$, where $x_{<t}$ denotes preceding tokens and $\mathcal{H}$ represents historical context, the steganographic process encodes secret bit stream $\mathbf{m} = (m_1, m_2, \dots, m_n)$ into generated text $T = \{x_0, \dots, x_m\}$. At each generation step, a subset $\mathbf{m}_t$ of the secret bit stream is mapped to token selection via a steganographic function $f$:
\begin{equation}
x_t = f\left(\mathbf{m}_t, \mathcal{K}, P(x_t \mid x_{<t}, \mathcal{H})\right),
\end{equation}
where $\mathcal{K}$ is a shared private key. To evade suspicion by third parties, advanced methods enforce consistency between the stego text distribution and the cover text distribution generated by the original language model.

The Adaptive Dynamic Grouping (\textit{ADG})~\cite{ADG} divides the vocabulary into groups $\{\mathcal{G}_i\}$ with equal probability masses $\sum_{x \in \mathcal{G}_i} P(x \mid x_{<t}, \mathcal{H}) = 2^{-k}$, enabling $k$ secret bit stream to be encoded per step through group selection. In contrast, the \textit{METEOR}~\cite{METEOR} algorithm converts secret bit stream $\mathbf{m}$ into a decimal value $\bar{B} \in [0, 1]$ using arithmetic coding. It selects tokens whose cumulative probability interval contains $\bar{B}$:
\begin{equation}
\bar{B} \in \left[{\sum}_{i=1}^{k-1} P(x_i \mid x_{<t}, \mathcal{H}), {\sum}_{i=1}^k P(x_i \mid x_{<t}, \mathcal{H})\right).
\end{equation}
The \textit{Discop}~\cite{Discop} method generates multiple distribution copies $\{P^{(j)}\}$ by applying interval offsets to the original distribution $P(x \mid x_{<t}, \mathcal{H})$, encoding secret bit stream through the index $j$ of the selected copy. These methods ensure covert communication while maintaining text quality and achieving computational indistinguishability between stego text and cover text.

However, modern large language models typically generate text with low-entropy characteristics, leading to limited payload capacity. This phenomenon in LLM-generated text originates from the sharp probability distributions of token predictions. This results in sparse high-probability tokens and restricts the available candidates for steganographic encoding. Consequently, existing methods require multiple rounds of communication to transmit secret messages, thereby increasing the risk of suspicion due to unusual interaction patterns. Our solution addresses this fundamental limitation by dynamically expanding the encoding space through entropy-aware token ranking, enabling efficient single-round secret transmission under practical constraints.

\section{Methodology}
\subsection{Overview.}
In this section, we provide a comprehensive overview of our proposed \textit{RTMStega} framework. Our framework comprises four core components: message encoding, message embedding, message extraction, and message decoding. As shown in Fig.~\ref{fig:framework}, \textit{RTMStega} integrates large language models to achieve lossless secret message compression and embedding through the entropy of the token probability distributions. In the embedding phase, secret messages are transcoded into compact rank sequences via LLM token predictions guided by a shared private context. This process leverages the inherent probability distributions of tokens to establish reversible mappings between secret bit stream and token ranks. Then, the rank sequences are encoded into generated text through entropy-constrained sampling with stego context. As for the extraction phase, recipients extract rank sequences from stego text transmitted via public channels and then reconstruct the original message through rank-guided token prediction using the shared context. The framework specifically addresses low-capacity limitations in stego text through multi-criteria ranking optimization. By expanding the effective encoding space via rank-token mapping based on the powerful LLM predictions, \textit{RTMStega} achieves higher embedding density than traditional grouping or arithmetic coding methods while ensuring text quality through entropy-driven sampling selection. 

We then introduce the details of each component as follows:

\subsection{Message Encoding.}
The proposed encoding framework transforms secret messages into compressed binary sequences using an autoregressive language model $\mathcal{M}$. This process combines token ranking prediction with pseudo-randomization to ensure both compression efficiency and stego text quality.

\noindent
\textbf{Token2Rank Conversion.} For a secret message $\mathbf{m} = (m_1, m_2, \dots)$, the encoder first tokenizes it into $\mathbf{X} = (x_1, x_2, \dots)$ using $\mathcal{M}$'s vocabulary $\mathcal{V}$. At each position $t \in |\mathbf{X}|$, the model computes the conditional probability distribution $p_\theta(x_t|x_{<t},\mathcal{H}_p)$ through forward propagation, where $\mathcal{H}_p$ denotes the shared private context. The vocabulary is sorted by descending probability to establish sorted sequences $\mathbf{r}_t$:
\begin{equation}
    \mathbf{r}_t = \text{argsort}(p_\theta(x_t|x_{<t},\mathcal{H}_p)).
\end{equation}
The ground-truth token's position $r_t$ in this sorted sequence is determined by:
\begin{equation}
    r_t = \text{index}(\mathbf{r}_t, x_t) \quad \text{where} \quad r_t \in \mathbb{Z}^+.
\end{equation}
After iteration, we can convert the secret message $\mathbf{X}$ into the corresponding rank sequence $\mathbf{R} = \{r_1, r_2, \dots\}$.

\noindent
\textbf{Binary Encoding \& Security Enhancement.} The rank sequence undergoes Huffman-inspired binary mapping $h(r_t)$, generating compressed bit stream $\mathbf{c} = \bigcup_{t=1}^{|\mathbf{R}|} h(r_t)$, where $\cup$ denotes bit stream concatenation. To enhance steganographic security, we implement cryptographic pseudo-randomization using a Pseudo-Random Number Generator (PRNG). The PRNG with the private key $\mathcal{K}$ produces a keystream matching the length of the message, which is XOR-operated with $\mathbf{c}$ to create the final cipher text $\mathbf{c}' = \{c'_1, c'_2, \cdots\}$. This dual-stage process ensures both information-theoretic efficiency and cryptographic randomness in the encoded output.

\begin{algorithm}[t]
\caption{Stego Text Generation.}
\label{algo1}
\SetAlgoLined
\SetKwInOut{Input}{Input}
\SetKwInOut{Output}{Output}
\Input{$\mathcal{M}$, $\mathcal{V}$, $\mathcal{K}$, $\mathbf{m}$, $\mathcal{H}_p$, $\mathcal{H}_s$, $h(\cdot)$.}
\Output{$\mathbf{T}_{\text{stego}}$}
\BlankLine
$\mathbf{X} \leftarrow \mathcal{V}(\mathbf{m}), \; \mathbf{c} \leftarrow \emptyset$\;
\For{$t \leftarrow 1$ \textbf{to} $|\mathbf{X}|$}{
    $r_t \leftarrow \operatorname{token2rank}(x_{<t},x_t,\mathcal{H}_p)$\;
    $\mathbf{c} \leftarrow \mathbf{c} \cup h(r_t)$\;
}
$\mathbf{c}' \leftarrow \mathbf{c} \oplus \text{PRNG}(\mathcal{K}, |\mathbf{c}|)$ \; 
$\mathbf{d} \leftarrow \operatorname{\beta-bit}(\mathbf{c}')$\;
$\mathbf{T} \leftarrow \emptyset, i \leftarrow 1$\;
\For{$t \leftarrow 1$ \textbf{to} $|\mathbf{d}|$}{
    $p_\theta^i \leftarrow \mathcal{M}(\mathbf{T},\mathcal{H}_s)$\;
    $E_i \leftarrow \operatorname{norm-entropy}(p^i_\theta)$\;
    \If{$E_i \geq \alpha \cdot \beta$}{
        $x_i \leftarrow \operatorname{rank-sampling}(d_t,\mathcal{H}_s)$\;
    }
    \Else{
        $x_i \leftarrow \operatorname{random-sampling}(\mathcal{H}_s)$\;
        $t \leftarrow t-1$\;
    }
    $\mathbf{T} \leftarrow \mathbf{T} \cup \{x_i\}$\;
}
$\mathbf{T}_{\text{stego}} \leftarrow \mathcal{V}(\mathbf{T})$ \;
\Return{$\mathbf{T}_{\text{stego}}$}\;
\label{algo1}
\end{algorithm}

\noindent
\textbf{$\beta$-bit Transformation.} We note that even for the same rank sequence, the text content obtained by decompression in different contextual environments is completely different. We use this property for the next message embedding. However, considering the uncertainty of the entropy of the embedding process, we need to transform the obtained pseudo-random bit stream $\mathbf{c}'$ into $\beta$-bit rank sequences $\mathbf{d} = \{d_1, d_2, \cdots\}$:
\begin{equation}
    d_t=\sum_{i=1}^{\beta}c'_{j} \cdot 2^{\beta-i}, \quad j= (t-1)\cdot \beta+i.
\end{equation}
The $\beta$ parameter balances the payload capacity and stego text quality. Larger $\beta$ values enable higher payloads but increase decoding uncertainty, while smaller $\beta$ enhances text quality at reduced capacity. This transformation ensures the embedded ranks align with the target language model's probability distribution characteristics.

\subsection{Message Embedding.}
The embedding process embeds the $\beta$-bit rank sequence $\mathbf{d}$ into text generation through entropy-constrained adaptive sampling.

\noindent
\textbf{Calculate Norm-Entropy.} At each generation step $t$ with stego context $\mathcal{H}_s$, we first extract the top-$2^\beta$ candidate tokens $A_t = \{a^1_t, \dots,a^{2^\beta}_t\}$ sorted by descending probability. Their probabilities are normalized as:
\begin{equation}
    \tilde{p}_t^i = \frac{p_\theta(a_t^i|x_{<t},\mathcal{H}_s)}{\sum_{j=1}^{2^\beta} p_\theta(a_t^j|x_{<t},\mathcal{H}_s)}.
\end{equation}
Then we use threshold $\alpha \in (0,1)$ controls sampling process via entropy constraint:
\begin{equation}
    E_t = -\sum_{i=1}^{2^\beta} \tilde{p}_t^i \log_2 \tilde{p}_t^i \geq \alpha \cdot \beta.
\end{equation}

\noindent
\textbf{Rank Sampling.} When $E_t \geq \alpha \cdot \beta$, we select token $a_t^{d_t}$ corresponding to rank $d_t$, which embeds the $\beta$-bit information. For low-entropy cases ($E_t < \alpha \cdot \beta$), we perform normal random sampling without embedding to maintain text quality. The final stego text $\mathbf{T}_\text{stego}$ is obtained by decoding the token sequence $\mathbf{T}$ through vocabulary $\mathcal{V}$, achieving covert communication through public channels. The entire generation process, from the secret message to the stego text, is shown in Algorithm 1.

\subsection{Message Extraction.}
The extraction process reconstructs secret messages from stego text through inverse operations of the encoding framework, requiring the identical language model $\mathcal{M}$, stego context $\mathcal{H}_s$, and the private key $\mathcal{K}$ for cryptographic consistency.

\noindent
\textbf{Candidate Set Reconstruction.} For each token $x_t$ in tokenized stego text $\mathbf{T}$, the receiver reconstructs the candidate set $A_t$ and normalized probabilities $\tilde{p}_t^i$ using the same entropy calculation as the embedding phase. The secret bit stream recovery follows:
\begin{equation}
\begin{aligned}
&s_t = 
\begin{cases}
\text{bin}(\tilde{r}_t) & \text{if } E_t \geq \alpha \cdot \beta \\
\epsilon & \text{otherwise}
\end{cases} \\
\quad &\text{where} \quad \tilde{r}_t = \sum_{k=1}^{|A_t|} \mathbb{I}(p(x_t) \geq \tilde{p}_t^k).
\end{aligned}
\end{equation}
Here, $\mathbb{I}(\cdot)$ denotes the indicator function identifying positions where $x_t$'s probability exceeds candidate tokens.

\noindent
\textbf{Bit stream Reconstruction \& Decoding.} The concatenated bit stream $\mathbf{c}'$ undergoes cryptographic de-randomization using the same PRNG with $\mathcal{K}$, reversing the XOR operation from encoding. The Huffman codebook then decodes the de-randomized bit stream to rank sequence $\hat{\mathbf{r}}$, completing the closed-loop extraction.

\begin{algorithm}[t]
\caption{Secret Message Extraction.}
\label{algo2}
\SetAlgoLined
\SetKwInOut{Input}{Input}
\SetKwInOut{Output}{Output}
\Input{$\mathcal{M}$, $\mathcal{V}$, $\mathcal{K}$, $\mathbf{T}_{\text{stego}}$, $\mathcal{H}_p$, $\mathcal{H}_s$, $\overline{h}(\cdot)$.}
\Output{$\hat{\mathbf{m}}$}
\BlankLine
$\mathbf{T} \leftarrow \mathcal{V}(\mathbf{T}_\text{stego}), \; \mathbf{s} \leftarrow \emptyset$\;
\For{$t \leftarrow 1$ \textbf{to} $\left|\mathbf{T}\right|$}{
    $ s_t \leftarrow \operatorname{extract-secret-bits}\left(\hat{x}_t, \mathcal{H}_s\right)$\;
    $\mathbf{s} \leftarrow \mathbf{s} \cup s_t$ \;
}
$\mathbf{c}' \leftarrow \mathbf{s} \oplus \text{PRNG}(\mathcal{K}, |\mathbf{s}|)$\;
$\hat{\mathbf{r}} \leftarrow \overline{h}(\mathbf{c}'), \; \hat{\mathbf{X}} \leftarrow \emptyset$\;
\For{$t \leftarrow 1$ \textbf{to} $\hat{\mathbf{r}}$}{
    $\hat{x} \leftarrow \operatorname{rank2token}(\hat{r}_t,\mathcal{H}_p)$\;
    $\hat{\mathbf{X}} \leftarrow \hat{\mathbf{X}} \cup \hat{x}$\;
}
$\hat{\mathbf{m}} \leftarrow \mathcal{V}(\hat{\mathbf{X}})$ \;
\Return{$\hat{\mathbf{m}}$\;}
\end{algorithm}

\subsection{Message Decoding.}
The decoding process inversely transforms rank sequences back to original messages through coordinated autoregressive reconstruction with the sender's private context $\mathcal{H}_p$.

\noindent
\textbf{Inverse Rank2Token Mapping.} For each decoded rank $\hat{r}_t$, the receiver reconstructs the token sequence using the shared private context $\mathcal{H}_p$. At position $t$, the model regenerates the sorted index vector through forward computation:
\begin{equation}
    \mathbf{r}_t = \text{argsort}(p_\theta(\hat{x}_t|\hat{x}_{<t},\mathcal{H}_p)).
\end{equation}
The original token is recovered via direct rank indexing:
\begin{equation}
    \hat{x}_t = \mathbf{r}_t[\hat{r}_t].
\end{equation}

\noindent
\textbf{Autoregressive Reconstruction.} The decoding progresses autoregressively, where each reconstructed token $\hat{x}_t$ conditions subsequent predictions. The process terminates upon end-of-sequence detection, yielding the secret message $\mathbf{m}$. This closed-loop reconstruction guarantees perfect message recovery when:
\begin{itemize}
    \item \textbf{Contextual Alignment:} $\mathcal{H}_p$ and $\mathcal{H}_s$ matches between encoder/decoder;
    \item \textbf{Ranking Consistency:} Token sort order remains stable, i.e., stable sorting algorithms are needed;
    \item \textbf{Key Synchronization:} PRNG states are synchronized via $\mathcal{K}$.
\end{itemize}

The complete decoding workflow is formalized in Algorithm 2, establishing a transformation from the stego text to the secret message under identical cryptographic conditions.

\section{Experiment}
In this section, we establish a comprehensive evaluation framework encompassing experimental configuration, performance evaluation, parametric analysis, and security validation. In the next part, we first present the details of the experimental setup and then show the results in turn.
\begin{table*}[!t]
\centering
\caption{The primary performance metrics of \textit{RTMStega} are evaluated and compared with those of three mainstream steganography: \textit{ADG}~\cite{ADG}, \textit{METEOR}~\cite{METEOR}, and \textit{Discop}~\cite{Discop}, in terms of both efficiency and text quality.}
\label{tab:tab1}
\begin{tabular}{@{}cccccccc@{}}
\toprule
Model & Dataset & Method & Payload$(\%\uparrow)$ & Embedding Time$(sec\downarrow)$ & Extraction Time$(sec\downarrow)$ & PPL$(\downarrow)$ & $\text{PPL}_\text{20}$$(\downarrow)$ \\ \midrule
\multirow{15}{*}{Qwen2.5-7B} & \multirow{5}{*}{AGNews} & ADG & 4.52 & 60.94 & 60.52 & 5.97 & 15.91 \\
 &  & METEOR & 3.73 & 60.15 & 60.38 & 5.75 & 16.30 \\
 &  & Discop & 5.35 & 51.28 & 51.07 & 5.63 & 15.31 \\
 & & \cellcolor{lightgray} RTMStega & \cellcolor{lightgray}\textbf{17.28} & \cellcolor{lightgray}\textbf{21.66} & \cellcolor{lightgray}\textbf{21.60} & \cellcolor{lightgray}5.51 & \cellcolor{lightgray}15.63 \\
 &  & Random & \textbackslash{} & 11.72 & \textbackslash{} & 5.61 & 14.75 \\ \cmidrule(l){2-8} 
 & \multirow{5}{*}{IMDb} & ADG & 5.37 & 55.47 & 55.28 & 5.75 & 16.30 \\
 &  & METEOR & 4.06 & 61.28 & 61.26 & 5.70 & 16.06 \\
 &  & Discop & 5.91 & 45.41 & 45.24 & 5.67 & 14.30 \\
 & & \cellcolor{lightgray} RTMStega & \cellcolor{lightgray}\textbf{15.13} & \cellcolor{lightgray}\textbf{21.59} & \cellcolor{lightgray}\textbf{21.36} & \cellcolor{lightgray}5.58 & \cellcolor{lightgray}14.66 \\
 &  & Random & \textbackslash{} & 11.72 & \textbackslash{} & 5.61 & 14.75 \\ \cmidrule(l){2-8} 
 & \multirow{5}{*}{WikiQA} & ADG & 4.36 & 53.10 & 52.73 & 6.11 & 16.64 \\
 &  & METEOR & 4.64 & 60.71 & 60.24 & 6.82 & 16.32 \\
 &  & Discop & 5.08 & 43.41 & 43.62 & 5.68 & 16.03 \\
 & & \cellcolor{lightgray} RTMStega & \cellcolor{lightgray}\textbf{16.19} & \cellcolor{lightgray}\textbf{17.27} & \cellcolor{lightgray}\textbf{17.24} & \cellcolor{lightgray}5.81 & \cellcolor{lightgray}14.91 \\
 &  & Random & \textbackslash{} & 11.72 & \textbackslash{} & 5.61 & 14.75 \\ \midrule
\multirow{15}{*}{\begin{tabular}[c]{@{}c@{}}DeepSeek-R1-Distill\\ -Llama-8B\end{tabular}
} & \multirow{5}{*}{AGNews} & ADG & 3.31 & 69.62 & 69.50 & 6.24 & 19.26 \\
 &  & METEOR & 3.38 & 64.65 & 64.34 & 6.95 & 21.64 \\
 &  & Discop & 3.38 & 61.75 & 62.07 & 6.19 & 19.54 \\
 & & \cellcolor{lightgray} RTMStega & \cellcolor{lightgray}\textbf{10.67} & \cellcolor{lightgray}\textbf{30.40} & \cellcolor{lightgray}\textbf{30.24} & \cellcolor{lightgray}5.98 & \cellcolor{lightgray}21.68 \\
 &  & Random & \textbackslash{} & 11.74 & \textbackslash{} & 6.23 & 19.27 \\ \cmidrule(l){2-8} 
 & \multirow{5}{*}{IMDb} & ADG & 3.28 & 67.36 & 67.57 & 6.43 & 22.17 \\
 &  & METEOR & 3.02 & 63.82 & 64.18 & 6.99 & 21.15 \\
 &  & Discop & 3.82 & 64.93 & 64.51 & 6.67 & 19.03 \\
 & & \cellcolor{lightgray} RTMStega & \cellcolor{lightgray}\textbf{9.72} & \cellcolor{lightgray}\textbf{25.11} & \cellcolor{lightgray}\textbf{25.06} & \cellcolor{lightgray}6.41 & \cellcolor{lightgray}21.22 \\
 &  & Random & \textbackslash{} & 11.74 & \textbackslash{} & 6.23 & 19.27 \\ \cmidrule(l){2-8} 
 & \multirow{5}{*}{WikiQA} & ADG & 3.19 & 67.53 & 67.59 & 6.42 & 22.10 \\
 &  & METEOR & 3.11 & 63.97 & 63.85 & 5.99 & 22.32 \\
 &  & Discop & 3.73 & 56.31 & 56.21 & 6.19 & 20.21 \\
 & & \cellcolor{lightgray} RTMStega & \cellcolor{lightgray}\textbf{10.00} & \cellcolor{lightgray}\textbf{29.19} & \cellcolor{lightgray}\textbf{28.96} & \cellcolor{lightgray}6.38 & \cellcolor{lightgray}21.79 \\
 &  & Random & \textbackslash{} & 11.74 & \textbackslash{} & 6.23 & 19.27 \\ \bottomrule
\end{tabular}
\vspace{-0.4cm}
\end{table*}
\subsection{Experimental Setup.}
We present implementation specifics through four key aspects: model selection, dataset configuration, baseline parameters, and evaluation metrics.

\noindent
\textbf{Large Language Models.} Our experiments employ two mainstream open-source LLMs: Qwen2.5-7B~\cite{qwen2.5} and DeepSeek-R1-Distill-Llama-8B~\cite{2025deepseekr1}. For stego text generation, we implement random sampling with temperature parameter $\tau=0.7$, intentionally disabling top-$p$ and top-$k$ truncation methods to maintain sampling diversity, i.e., high sampling distribution entropy.

\noindent
\textbf{Dataset Configuration.} We utilize text datasets from AGNews (business category)~\cite{agnews}, IMDb~\cite{imdb}, and WikiQA~\cite{wikiqa} as secret messages. To address the issue of embedding lengthy secret messages that exceed the capacity of a single steganography session, we extract the first two sentences from each IMDb sample. For context consistency in general-purpose steganography, we establish a fixed prefix ``This is an emergency broadcast:'' as the shared private context. The Wikitext~\cite{wikitext} training set provides generation prompts, using the first two sentences of each entry as stego contexts.

\noindent
\textbf{Baseline Parameters.} Our proposed \textit{RTMStega} operates with hyperparameters $\alpha=0.6$ and $\beta=3$. The comparative analysis includes three mainstream generative steganography: \textit{ADG}~\cite{ADG}, \textit{METEOR}~\cite{METEOR}, and \textit{Discop}~\cite{Discop}. Implementation details strictly follow the original authors' configurations from respective open-source repositories. Additionally, given that baseline steganographic methods typically overlook the preprocessing of secret messages, we employ UTF-8 and Huffman encoding to preprocess and encode the secret messages.

\noindent
\textbf{Token Ambiguity:} Mainstream steganographic methods~\cite{ADG, METEOR, Discop} (including \textit{RTMStega}) require precise alignment of token paths between sender and receiver for successful decoding. This dependency becomes problematic in large language models (LLMs) like Qwen2.5 due to token ambiguity caused by byte-pair encoding (BPE)~\cite{bpe}. Specifically, BPE creates non-prefix-free vocabularies where multiple tokenization paths can represent the same input sequence. Alternative solutions, such as word-based tokenizers, character-based tokenization, or disambiguation algorithms~\cite{bauer2024leveraging,yan2023secure,qi2024provably} can reduce this ambiguity. To ensure fair performance comparison between our method and existing baselines, we implement a simplified token-passing mechanism that directly transmits stego tokens within the program workflow. This design choice eliminates interference from tokenization variations while maintaining fair comparisons between these steganographic methods.

\noindent
\textbf{Evaluation Metrics.} We evaluate \textit{RTMStega} from both efficiency and text quality perspectives: \textbf{Efficiency} metrics combine payload capacity (the ratio of the secret message length to the stego text length) and processing time. The processing time encompasses the average embedding time, defined as the duration from processing the secret message to generating the stego text, and the average extraction time, which involves extracting the bit stream from the stego text and reconstructing the secret message. \textbf{Text Quality} evaluation employs perplexity (PPL) for fluency assessment and contextual relevance analysis. The perplexity metric quantifies prediction confidence for text sequence $\mathbf{x} = \{x_1, \ldots, x_n\}$:
\begin{equation}
    \operatorname{PPL}(\mathbf{x}) = \exp\left(\frac{1}{n}\sum_{i=1}^n -\log P(x_i|x_{<i})\right)
\end{equation}
where $P(x_i|x_{<i})$ represents the model's next-token prediction probability. We employ Qwen2.5-14B~\cite{qwen2.5} as the evaluation model. Contextual relevance is measured through $\text{PPL}_\text{20}$, calculated using the final 10 context tokens and initial 10 generated tokens.

\subsection{Main Performance.}
The experimental results, as shown in Tab.~\ref{tab:tab1}, demonstrate significant performance advantages of \textit{RTMStega} over baseline methods in efficiency while maintaining text quality. This section provides a multi-dimensional analysis of the experimental performance and its underlying mechanisms.

\noindent
\textbf{Payload Capacity:}  \textit{RTMStega} achieves a 3× improvement in payload capacity compared to baseline methods across all model-dataset combinations. This stems from two key innovations: 1) The proposed dual-phase using token probability ranks as an intermediate representation, effectively increasing information density per token. 2) Dynamic probability threshold adjustment (controlled by parameters $\alpha$ and $\beta$) optimizes the trade-off between payload capacity and text naturalness, enabling higher embedding rates without triggering abnormal token selection.

\noindent
\textbf{Processing Time:} The \textit{RTMStega} framework achieves a remarkable reduction in both embedding and extraction times, outperforming baseline methods by more than 50\% per sample. This significant acceleration is primarily attributed to the increased payload capacity, which substantially decreases the number of tokens required to fully embed a secret message. This relationship is further supported by the observation that payload capacity is directly proportional to processing time.

\begin{figure*}[t]
    \centering
    \begin{subfigure}[b]{0.32\textwidth}
        \includegraphics[width=\textwidth]{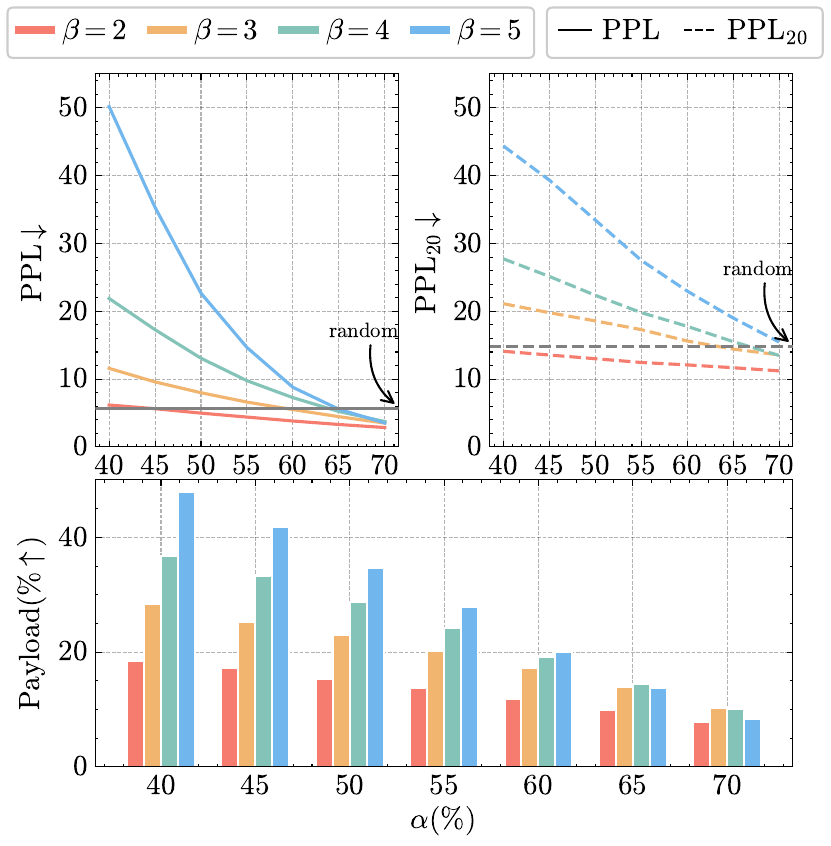}
        \caption{Qwen2.5-7B-AGNews}
        \label{fig:ag_news}
    \end{subfigure}
    \hfill
    \begin{subfigure}[b]{0.32\textwidth}
        \includegraphics[width=\textwidth]{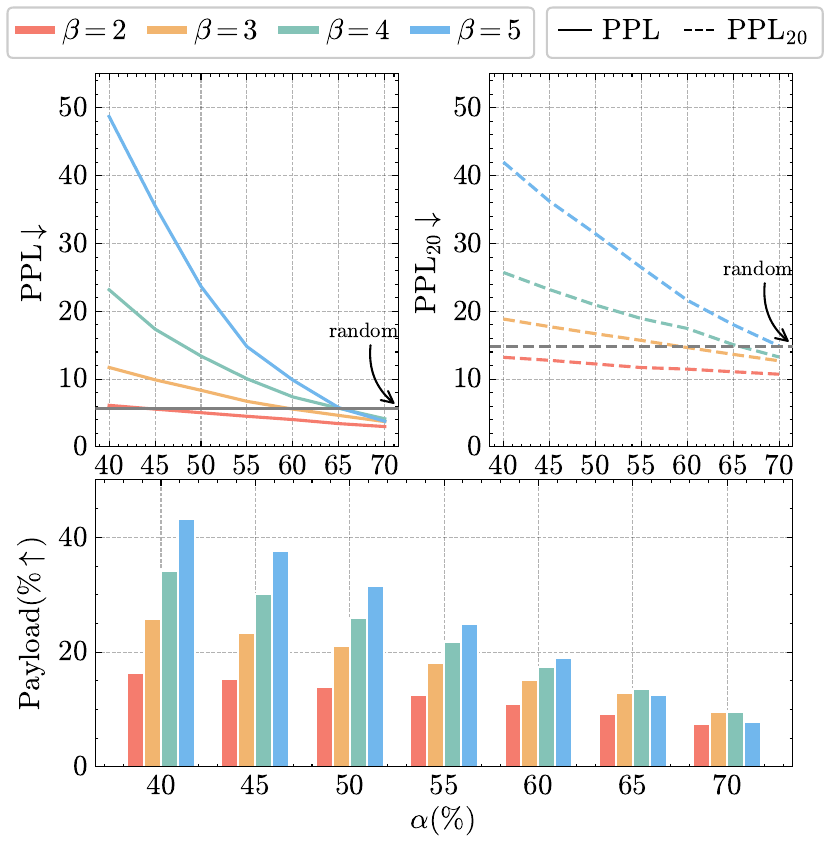}
        \caption{Qwen2.5-7B-IMDb}
        \label{fig:imdb}
    \end{subfigure}
    \hfill
    \begin{subfigure}[b]{0.32\textwidth}
        \includegraphics[width=\textwidth]{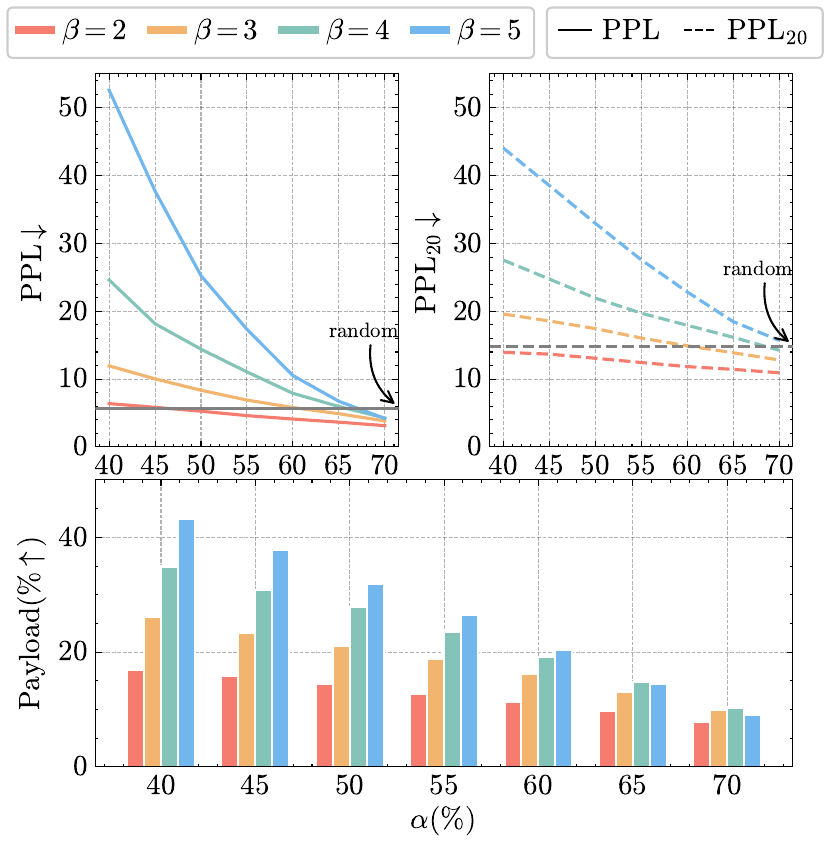}
        \caption{Qwen2.5-7B-WikiQA}
        \label{fig:wikiqa}
    \end{subfigure}
    \caption{Variation curves illustrating the relationship between stego text payload capacity and text quality as functions of $\alpha$ and $\beta$ are generated by \textit{RTMStega}. Here, $\alpha$ represents the threshold for determining whether to perform rank-based sampling based on the entropy of the model's output distribution, and $\beta$ dictates the volume of secret information embedded in a single step.}
    \label{fig:alpha_beta}
\vspace{-0.4cm}
\end{figure*}

\noindent
\textbf{Text Quality:} From the calculation results of PPL and $\text{PPL}_\text{20}$, the quality and correlation of text generated by \textit{RTMStega} are comparable to those of the baseline methods. This finding suggests that it is challenging to distinguish between stego text and cover text generated by \textit{RTMStega} based solely on text quality, thereby ensuring the security of our method. In addition, $\text{PPL}_\text{20}$ is usually much higher than the PPL since $\text{PPL}_\text{20}$ only calculates the average perplexity for a localized window at the intersection of prompt and stego text, where the model transitions from generating a prompt to generating new content, and context switching can make prediction more difficult, thus pushing up the perplexity level.

\noindent
\textbf{Model-Dependent Variance:} The cryptographic load rates on the DeepSeek-R1-Distill-Llama-8B model are generally lower than those on the Qwen2.5-7B model due to the fact that the DeepSeek-R1-Distill-Llama-8B model performs better in generating text than the Qwen2.5-7B model. As a result, the output has a lower entropy of the sampling distribution and fewer secret bit streams embedded in a single token. This is a generic problem for all methods of generating stego text by modifying the output probabilities. In addition, the lower PPL and $\text{PPL}_\text{20}$ calculated on the Qwen2.5-7B model is due to the fact that we use Qwen2.5-14B as the model for calculation, which has a similar architecture and a closer output distribution.

\subsection{Parameter Explorement.}
Based on the experimental results across AGNews, IMDb, and WikiQA datasets, we analyze the impact of parameters $\alpha$ (entropy threshold) and $\beta$ (correlates with the number of bit stream embedded in a single step) on steganographic performance. We choose Qwen2.5-7B as the language model, and the results are shown in Fig.~\ref{fig:alpha_beta}.

\noindent
\textbf{Analysis on $\alpha$.} Higher values of $\alpha$ strengthen the relevance of the generated text to the given context, forcing the model to prioritize the selection of words that are logically coherent with the context, which can effectively inhibit the degradation of the text quality but at the same time compresses the selection space of candidate words, making the information embedding efficiency tend to be conservative. When $\alpha$ is lowered, the model's dependence on context is weakened, allowing more unconventional but information-dense words to be selected, at which point the payload is significantly improved. However, the generated text may have segments that are out of context with the initial context, and this deviation is amplified through the cumulative effect of the lexical chain, which is reflected in the fact that the quality of the text decreases more rapidly as $\alpha$ is lowered. In addition, when $\alpha$ is high, the model tends to sample more randomly. However, due to the limitation of $\beta$, when the entropy of the sampling distribution of the model output is higher, the model will choose the candidate words with higher sampling probability, so the final overall text quality is better compared to random sampling results without top-$k$ and top-$p$ settings.

\noindent
\textbf{Analysis on $\beta$.} As $\beta$ increases, the model relaxes its filtering of candidate tokens, allowing more low-probability tokens to enter the embedding selection range, i.e., more secret information can be embedded in a single step, which directly enhances the payload capacity. However, when $\beta$ exceeds a certain range, the introduction of low-frequency tokens can disrupt the coherence of the text, leading to semantic breaks or syntactic anomalies in the generated content, which is manifested as a significant rise in the perplexity level (PPL and $\text{PPL}_\text{20}$). In addition, when $\beta$ is small, the candidate tokens of the model are only in the top-$2^\beta$ range, and the quality of the candidate tokens is all higher. Thus, the text quality is better relative to the random sampling results.

In this paper, we consider the need for security of steganographic behavior and choose the parameter configuration that is closest to the quality of the randomly sampled generated text, i.e., the combination of $\alpha=0.6$ and $\beta=3$.

\begin{table}[t]
\centering
\caption{Steganalysis accuracy for \textit{RTMStega}.}
\label{tab:tab2}
\setlength{\tabcolsep}{3pt}
\begin{tabular}{ccccc}
\toprule
Steganalyzer & TS-FCN$(\%)$ & LSTMATT$(\%)$ & BiLSTMDENSE$(\%)$ \\ \midrule
AGNews & 48.55$\pm$2.58 & 54.25$\pm$0.99 & 56.15$\pm$1.40 \\
IMDb & 49.15$\pm$1.58 & 54.85$\pm$1.74 & 55.90$\pm$3.27 \\
WikiQA & 48.85$\pm$2.60 & 55.10$\pm$3.06 & 57.15$\pm$0.64 \\ \bottomrule
\end{tabular}
\vspace{-0.4cm}
\end{table}

\subsection{Steganalytic Results.}
To assess the imperceptibility of \textit{RTMStega} in generating stego text, we conducted a comprehensive steganalysis experiment. Using the Qwen2.5-7B model, we generated two datasets: 1,000 cover texts and 1,000 stego texts. These datasets are analyzed using three linguistic steganalyzers: \textit{TS-FCN}~\cite{yang2019fast}, \textit{LSTMATT}~\cite{yu2020attention}, and \textit{BiLSTMDENSE}~\cite{yang2020linguistic}, all of which leverage pre-trained BERT embeddings~\cite{2019bert}. The datasets are partitioned into training, validation, and testing sets with a ratio of $3:1:1$. The experiments are configured with a learning rate of $1e-4$ and trained for five epochs. To ensure robustness, the process was repeated five times, and the average accuracy of the test set was used as the steganalysis accuracy metric.

The results, presented in Tab.~\ref{tab:tab2}, demonstrate that the detection error rates for \textit{RTMStega} across all three datasets are close to 50\%. This suggests that the steganalyzers perform no better than random guessing when attempting to differentiate between cover text and stego text. These findings strongly validate the security of \textit{RTMStega}, as its stego text effectively resists detection by mainstream linguistic steganalysis tools.

\section{Conclusion}
This paper introduces \textit{RTMStega}, a linguistic steganography framework that addresses the low-payload bottleneck in generative steganography through entropy-driven rank-token mapping. By harnessing the compression capabilities of large language models and implementing context-adaptive entropy normalization, \textit{RTMStega} encodes secret messages into ranked token sequences while maintaining the naturalness and statistical consistency of the stego text. Experimental results confirm that the proposed method significantly surpasses existing steganography, achieving a triple improvement in payload capacity, a 50\% reduction in processing time, and maintaining comparable text quality across multiple datasets and models. These advancements significantly enhance the practicality of covert communication systems by enabling higher throughput without compromising security. However, despite these achievements, the security of \textit{RTMStega} has not yet been provably established compared to baseline methods. Future research will aim to explore adaptive entropy thresholds to further optimize the balance between payload capacity and text quality while enhancing its undetectability.

\vspace{12pt}

\end{document}